\documentstyle[aps,prl,epsfig,multicol]{revtex}
\input epsf

\begin{document}
\draft

\title{Remnant Fermi surface in the presence of an underlying 
instability in layered 1$T$-TaS$_{2}$}
\author{Th. Pillo$^{(1)}$, J. Hayoz$^{(1)}$, H. Berger$^{(2)}$, M. 
Grioni$^{(2)}$, L. 
Schlapbach$^{(1)}$, P. Aebi$^{(1)}$}
\address{$^{(1)}$Institut de Physique, Universit\'e de Fribourg, CH-1700 
Fribourg, Switzerland\\ 
$^{(2)}$ Institut de Physique Appliqu\'ee, EPF, CH-1015 Lausanne, 
Switzerland\\
\vspace{0.5cm}\textnormal {submitted to Phys. Rev. Lett., \today}}
\maketitle
\begin{abstract}      
We report high resolution angle-scanned photoemission and Fermi 
surface (FS) mapping experiments on the layered 
transition-metal dichalcogenide 1$T$-TaS$_{2}$ in the quasi 
commensurate (QC) and the commensurate (C) charge-density-wave (CDW) phase. 
Instead of a nesting induced partially removed FS in the CDW phase we 
find a pseudogap over large portions of the FS. This remnant FS 
exhibits the symmetry of the one-particle normal state FS even when 
passing from the QC-phase to the C-phase. Possibly, this Mott localization 
induced transition represents the underlying instability 
responsible for the pseudogapped FS.
\end{abstract} 

\begin{multicols}{2}
\narrowtext	
The layered transition metal dichalcogenide 1$T$-TaS$_{2}$ is a model system 
being the first material where charge density waves (CDW) have been 
experimentally discovered by means of superlattice spots in X-ray 
diffraction (XRD) experiments\cite{wilson74}. It provides, however, one crucial 
difference to isostructural 1$T$-type materials, because it shows, 
besides the Peierls transition at approximately 550~K, a Mott localization 
induced transition at 
180~K, where electrons stemming from the Ta~$5d$-band manifold become 
more and more localized with decreasing temperature and, suddenly, 
yield a commensurate locked-in CDW\cite{fazekas80}. As a 
consequence, 1$T$-TaS$_{2}$ exhibits a rich phase diagram, 
where several phases exist as a function of 
temperature\cite{wilson75}. The 
relevant phases here are, first,  the quasi-commensurate (QC) phase, 
stable at room temperature (RT) and known to exhibit 
hexagonal arrays of commensurate domains with 
$(\sqrt{13} \times \sqrt{13})$ symmetry\cite{stm}. Second, the commensurate (C) 
phase below 180~K where the CDW is completely locked in. 
The electronic structure of 1$T$-TaS$_{2}$ is considerably influenced 
by the CDW expressed by the decay of the one-particle Ta~$5d$-band, which is 
split off into three dispersionless submanifolds already in the 
QC-phase\cite{smith85,manzke88}. Angle-resolved photoelectron 
spectroscopy (ARPES) investigations have given experimental 
evidence\cite{manzke89,dardel92a,dardel92b} 
for a so-called $'$star-of-David$'$ model of Fazekas and Tosatti 
(FT)\cite{fazekas80}. Thirteen Ta atoms form two outerlying bonding shells with six Ta atoms 
each, displayed in ARPES spectra as the two low energetic manifolds. 
The shallow band containing the remaining thirteenth electron 
is susceptible to Mott localization and splits into a lower occupied 
(LHB) and upper unoccupied Hubbard subband. The LHB is manifest as a sharp, dispersionless 
peak near the Fermi level E$_{F}$ in near-normal emission ARPES 
spectra of the C phase\cite{manzke89,dardel92a,dardel92b}. Complementary 
tunneling spectroscopy data\cite{kim94} indicates a 
symmetric  splitting of the LHB and the unoccupied upper band with respect 
to E$_{F}$. Temperature 
dependent near normal emission measurements showed, that the C-phase reveals a pseudogap 
with residual spectral weight at E$_{F}$ down to very low temperatures, 
explaining low temperature resistivity data in terms of a variable range 
hopping mechanism\cite{dardel92a,dardel92b}.
Moreover, the local Coulomb correlation energy, the Hubbard U$_{dd}$, depends on random 
disorder\cite{grioni98}. 

Very recently Fermi surface (FS) measurements using ARPES have gained particular 
interest with respect to the mechanism behind high-temperature 
superconductivity\cite{ronning98,norman98}. In underdoped cuprates a remnant FS has been 
detected at temperatures around the transition 
temperature T$_{c}$\cite{norman98}. The important point 
seems to be 
that there is an underlying electronic instability which drives 
the pseudogap and the remnant FS behaviour\cite{ding97,ding96}. One might ask whether to expect a 
comparable behaviour of the FS for other materials with underlying 
electronic phase transitions. 
As a possible candidate appears 1$T$-TaS$_{2}$, which has as an 
underlying instability the Mott localization derived transition at 180~K, where no new symmetry is 
dictacted\cite{fazekas80}. Above the transition temperature, in the 
QC phase an intact FS should be present except for new zone 
boundaries where a (Peierls) gap is opened\cite{gruener}.

In the present Letter we show, using temperature dependent 
\textit{scanned} ARPES and FS mapping (FSM) measurements on 1$T$-TaS$_{2}$ that the FS is 
(pseudo)gapped although only a partial removal of the FS due 
to nesting is expected. Actually, it is a remnant FS yielding the symmetry of the 
normal (metallic) state (NS) even below 180~K with residual spectral weight at E$_{F}$.

Experiments have been performed in a modified VG ESCALAB Mk II 
spectrometer using monochromatized He-I$\alpha$ (21.2 eV) 
photons\cite{pillo98}.The FSM measuring mode including sequential 
motorized sample rotation has been outlined previously\cite{fsm}. Note that this type 
of \textit{scanned} ARPES measurement yields \textit{direct} information about relevant points 
in \textbf{k}-space, in contrast to traditional ARPES work, where the datasets 
are intrinsically smaller and possibly omit information. 
ARPES and FSM experiments were performed with energy and angular resolution of 
$30$ meV and $\pm 0.5^{\circ}$, respectively. 
Pure 1$T$-TaS$_{2}$ samples were prepared by standard flux growing 
techniques\cite{dardel92a,dardel92b} and cleaved 
\textit{in situ} at pressures in the upper $10^{-11}$ mbar region. Surface quality and 
cleanness have been checked by low energy electron diffraction (LEED) and X-ray 
photoelectron spectroscopy, respectively. X-ray photoelectron diffraction\cite{osterwalder} allows us 
to determine the sample orientation \textit{in situ} with an accuracy 
of better than 0.5$^\circ$.

Figure 1 shows FSM data of  1$T$-TaS$_{2}$ taken with monochromatized He-I$\alpha$ 
radiation (21.2~eV) in the QC-phase (a) and in the C-phase (b). The (1x1), i.e., NS-surface Brillouin 
zones (SBZ) are given by the hexagons with the corresponding high symmetry points 
indicated. Additionally, in panel (b) the 
(expected) new SBZ due to the commensurate CDW is illustrated by the 
small hexagons. The raw data has been normalized\cite{norm} such that the overall polar intensity variation has 
been removed and symmetrized according to the space group 
D$_{3d}^3$\cite{wilson75} of the CdI$_{2}$-type materials. In all panels the measurements 
are given in a parallel-projected linear gray scale representation with maximum intensity 
corresponding to white. The center of the plots denotes normal emission ($\Gamma$, polar 
emission angle $\theta=0^\circ$) and the outer circle represents grazing emission 
($\theta=90^\circ$). Elliptic branches, centered around M, as expected from band-structure 
calculations\cite{calc} are seen. A comparable shape has been found 
in a dichroism study of the valence band of 1$T$-TaS$_{2}$\cite{matsushita}
When comparing the data at 295~K and 140~K, two features are surprising. First, at 295~K 
as well as at 140~K, the symmetry is threefold and not broken 
according to the $(\sqrt{13} \times \sqrt{13})$ reconstruction as indicated by the small 
hexagons in panel (b). This might be explained by a small Fourier 
component of the CDW potential\cite{claessen90}, at least for 295~K. 
Nevertheless, this is surprising since 1$T$-TaS$_{2}$ is reported to 
have a very strong CDW amplitude\cite{wilson75,stm}. Second, for 140~K, one 
would not  expect to see any FS at all due to the reported rigid 
quasiparticle (QP) band shift of 180~meV in the C-phase\cite{manzke89}. 
Instead we observe small but finite spectal weight at E$_{F}$ all 
over the $''$FS$''$.

In order to get more detailed information about the actual behaviour of the FS, 
ARPES spectra were measured along the FS contour. Spectra are given in 
Fig.~2 for 295~K and 140~K, respectively. The arrow on the right side 
indicates the \textbf{k}-space location of the spectra 1--41 as 
displayed in Fig.~3 (bottom) where the NS-FS is 
mimicked by ellipses. 
To obtain the exact FS contour locations, we measured 
the spectral weight at E$_{F}$ within the BZ wedge $\Gamma$KMK$\Gamma$ (see Fig.~3) with a 
very high point density including approximately 5400 angular positions. 
For 295~K one observes two broad QP peaks (Fig.~2), denoted A and B in the 
spectra at energy positions of about 220~meV and~700 meV, 
respectively. Both peaks show significant modulation 
along the FS contour but practically no dispersion. Most importantly, there is no clear Fermi edge 
visible. Instead we observe a leading edge shift of at least 30~meV 
as compared ot the Fermi edge of the polycrystalline Cu sampleholder. 
In the low temperature C-phase spectra 
(right panel) along the FS contour the situation is very similar 
expect that both peaks yield a rigid shift of about 120 meV 
and become narrower. Despite the energy shift the gap does not open 
completely and the ARPES spectra retain small but finite spectral weight at E$_{F}$. 
Note that the same \textbf{k}-space dispersion is observed as in (a), 
however with the maximal intensity reduced by approximately an order 
of magnitude. This means that the QC-C transition does 
not change the symmetry of the $''$FS$''$ anymore. 
We interpret this in terms of a pseudogap due to  finite 
hybridization of the overlapping tails of the two Hubbard subbands. 
The pseudogap remains open all over the $''$FS$''$, indicating that 
there is no clear Fermi level crossing of a one-particle peak, even 
along the $\left[ \Gamma \mbox{M} \right]$ close to 
$\Gamma$, where the intensity is enhanced\cite{pillo99}. 
Instead one has a remnant Fermi surface (RFS) already at room 
temperature. Over all, this RFS has, as seen from Fig.~1(a) and (b) 
the symmetry retained from the elliptic  one-particle NS-FS. 
For convenience, Fig.~3 shows a sketch of the situation in 
1$T$-TaS$_{2}$. The upper part shows the E(\textbf{k}) of the one-particle NS Ta 5$d$ band 
with the Fermi level crossing in the normal phase at about 1/3 of the $\Gamma$M 
distance\cite{smith85,manzke88,grioni98,calc,pillo99}. The 
$''$shady$''$ areas indicate the practically dispersionless subbands 
due to the CDW potential splitting the Ta 5$d$ 
band\cite{fazekas80,smith85}. 
The analoguous \textbf{k$_{\|}$}(E$_{F}$) is plotted in the lower part as the NS-FS. In addition we superimposed on 
the NS-FS the Peierls gap region as a gray part on the ellipse around the 
small half-axis. Since the loss of intensity of peak B appears exactly in 
this region (Fig.~2), it may be attributed to removal of 
spectral weight due to the Peierls gap\cite{pillo99}. From Fig.~2 it is now clear, that the 
FS is disrupted to a RFS yet retaining the overall NS symmetry 
even below 180~K.

In order to corroborate the pseudogap along the RFS 
contour probed in Fig.~2, we measured 
azimuthal ARPES spectra along the two circular arcs shown in Fig.~3, 
i.e., along a$\rightarrow$b and c$\rightarrow$d. 
Results for RT are shown in Fig.~4 (a) and (b), respectively. Figure~4(a) 
shows the spectra for a polar angle of 32$^\circ$ (a$\rightarrow$b) thus 
crossing the 
Peierls gap region as indicated in Fig.~3. Figure 4(b) shows the spectra 
for a polar angle of 50$^\circ$ (c$\rightarrow$d), being outside the  
region influenced by the CDW. In both panels we display a 
gray scale representation and the spectra. The arrows indicate the 
Fermi level crossing of the NS band as obtained from plots of the 
occupation number n(\textbf{k}). In both panels a single dispersing band can be seen sitting on 
an incoherent background which is built from 
non-dispersing QP peaks, corresponding to the decay of the NS band into 
three submanifolds\cite{fazekas80}. At $\theta=32^\circ$ polar angle, the dispersing band 
reaches down to 0.95 eV binding energy in the $\left [\Gamma \mbox{M} \right ]$ 
azimuth, and for $\theta=50^\circ$ we have 0.5 eV as the maximum binding 
energy. The maximum binding energies fit well with the fact that the 
a$\rightarrow$b scan crosses at the full depth of the band near M (see 
Fig.~3) whereas the c$\rightarrow$d scan probes a more shallow part 
near the end of the ellipse. As a guide to the eye, the peak positions are shown, 
respectively, as white circles 
overplotted on the gray scale map and as small ticks in the spectra. 
Strikingly, one can see the backdispersing of the band when it reaches 
the region indicated by the arrows. However, there is \textit{no} QP band crossing 
the Fermi level anywhere in the irreducible BZ wedge confirming that 
the original FS is completely pseudogapped and remains a RFS.
These findings are remarkable insofar as one would 
expect a simple removal of FS portions due to nesting, either via large 
parallel areas (i.e., the gray shaded portion of the ellipse; bottom of 
Fig.÷3) of the NS-FS or via a 
very strong electron-phonon coupling constant. Other nested FS regions 
than those indicated by the gray zones in Fig.~3 are not expected 
because XRD\cite{wilson74} and LEED
data\cite{pillo99b} clearly show the existence of a nesting vector corresponding 
to the superimposed ($\sqrt{13}$x$\sqrt{13}$) symmetry (indicated by 
the small hexagons in Fig.~1(b)). Originally the lattice distortion may be driven by nesting, but 
the electronic structure is more and more 
influenced as the localization of star-centered Ta electrons sets in. 
This remaining thirteenth electron in the star, the LHB, becomes more and more 
localized and gives rise to the one order of magnitude increase of the 
in-plane resistivity at the QC-C transition\cite{fazekas80}. The QC-phase 
exhibits commensurate domains with an incoherent 
superposition of domains. Hence, the overlap of the QP 
peaks is considerable but decreases with temperature\cite{claessen90}. 
At the QC-C transition there is probably a sudden increase of 
U$_{dd}$ which pops the two Hubbard subbands apart. Interestingly, the 
remaining spectral weight distribution keeps the same symmetry. From 
Fig.~4 we note that at RT we are left with \textit{one} dispersing QP band 
whereas the other two bands are visible as shoulders in the QP peak, 
and already exhibit a dispersionless behaviour. Nonetheless we do not have 
a Fermi edge crossing. 
We may explain this when we consider the Mott 
transition at 180~K as an underlying transition. At RT 
the pseudogap might be interpreted as a kind of precursor. There is a striking similarity with 
underdoped cuprates\cite{norman98,ding97,ding96} where a crossover temperature 
scale T$^\star >$ T$_{c}$ has been introduced\cite{ding97}. Between T$^\star$ 
and T$_{c}$ a pseudogap opens leading to the complete opening of the 
superconducting gap at T$_{c}$. One is tempted to 
introduce such a T$^{\star}$ as well for 1$T$-TaS$_{2}$, 
i.e. at the onset of the localization where the CDW becomes 
quasicommensurate. These findings should also motivate a reconsideration of 
isoelectronic systems, where electron-phonon and electron-electron 
interactions do interfere. 

In summary, we have shown by means of temperature dependent scanned 
ARPES and FSM mesurements that at room temperature 1$T$-TaS$_{2}$ 
yields a remnant Fermi surface which is not only affected by a comparingly 
small influence of the CDW formation, i.e., nesting, but rather it is the underlying Mott 
localization induced transition which seems to tune FS properties. This remnant FS is 
retained even below the QC-C transition with the same symmetry than in 
the metallic normal state phase. As in underdoped cuprates, the introduction 
of a crossover temperature has been suggested. 

We would like to thank D. Baeriswyl and R. Noack for stimulating 
discussions. The outstanding help of our workshop and electronic team with O. 
Raetzo, E. Mooser, Ch. Neururer, and F. Bourqui is gratefully
acknowledged. This work has been supported by the Fonds National 
Suisse pour la Recherche Scientifique.

%%%%%%%%%%%%%%%%%%%%%%%%%%%%%%%%%%%%%%%%%%%%%%%%%%%%%%%%%%%%%%%%%%%%%%%%%
%\bibliographystyle{prsty}
%\bibliography{ups}

\begin{thebibliography}{10}
\bibitem{wilson74}
J.~A. Wilson, F.~J. Di~Salvo, S. Mahajan, Phys. Rev. Lett. \textbf{32}, 882 (1974).
\bibitem{fazekas80}
P. Fazekas, E. Tosatti, Phil. Mag. B \textbf{39}, 229 (1979); 
P. Fazekas, E. Tosatti, Physica \textbf{B 99}, 183 (1980).
\bibitem{wilson75}
J.~A. Wilson, F.~J. Di~Salvo, S. Mahajan, Adv. Phys. \textbf{24}, 117 (1975).
\bibitem{stm}
X. Wu et al., Science \textbf{243}, 1703 (1989); B. Giambattista 
et al., Phys. Rev. B \textbf{41}, 10082 (1990); W. Han et al.,
\textit{ibid} \textbf{48}, 8466 (1993); B. Burk et al., Phys. 
Rev. Lett. \textbf{66}, 3040 (1991); X. Wu et al., \textit{ibid} 
\textbf{64}, 1150 (1990).
\bibitem{smith85}
N.~V. Smith, S.~D. Kevan, F.~J. Di~Salvo, J. Phys. \textbf{C 18}, 3175 (1985).
\bibitem{manzke88}
R. Manzke, O. Anderson, M. Skibowski, J. Phys. \textbf{C 21}, 2399 (1988).
\bibitem{manzke89}
R. Manzke, T. Buslaps, B. Pfalzgraf, M. Skibowski, O. Anderson, Europhys. Lett. \textbf{8}, 195 (1989).
\bibitem{dardel92a}
B. Dardel, M. Grioni, D. Malterre, P. Weibel, Y. Baer, F. L\'evy, Phys. Rev. B \textbf{45}, 1462 (1992).
\bibitem{dardel92b}
B. Dardel, M. Grioni, D. Malterre, P. Weibel, Y. Baer, F. L\'evy, Phys. Rev. B \textbf{46}, 7407 (1992).
\bibitem{kim94}
J.~J. Kim, W. Yamaguchi, T. Hasegawa, K. Kitazawa, Phys. Rev. Lett. \textbf{73}, 2103 (1994).
\bibitem{grioni98}
F. Zwick, H. Berger, I. Vobornik, G. Margaritondo, L. Forr\'o, M. 
Onellion, G. Panaccione, A. Taleb, M. Grioni, Phys. Rev. Lett. \textbf{81}, 1058 (1998).
\bibitem{ronning98}
F. Ronning, C. Kim, D.~L. Feng, D.~S. Marshall, A.~G. Loeser, L.~L. 
Miller, J.~N. Eckstein, I. Bozovic, Z.-X. Shen, Science \textbf{282}, 2067 (1998).
\bibitem{norman98}
M.~R. Norman, H. Ding, M. Randeria, J.~C. Campuzano, T. Yokoya, T. 
Takeuchi, T. Takahashi, T. Mochiku, K. Kadowaki, P. Guptasarma, D.~G. 
Hinks, Nature (London) \textbf{392}, 158 (1998).
\bibitem{ding97}
H. Ding, M.~R. Norman, T. Yokoya, T. Takeuchi, M. Randeria, J.~C. 
Campuzano, T. Takahashi, T. Mochiku, K. Kadowaki,  Phys. Rev. Lett. \textbf{78}, 2628 (1997).
\bibitem{ding96}
H. Ding, T. Yokoya, J.~C. Campuzano, T. Takahashi, M. Randeria, M.~R. 
Norman, T. Mochiku, K. Kadowaki, J. Giapintzakis,  Nature (London) \textbf{382}, 51 (1996).
\bibitem{gruener}
G. Gr\"{u}ner, \textit{Density Waves in Solids}, Addison-Wesley 
(1994) and references therein.
%\bibitem{boschung98}
%E. Boschung, Th. Pillo, J. Hayoz, L. Patthey, P. Aebi, L. Schlapbach, Phys Rev. B \textbf{58}, 10210 (1998).
\bibitem{pillo98}
Th. Pillo, L. Patthey, E. Boschung, J. Hayoz, P. Aebi, L. Schlapbach, J. Electron Spectr. Relat. Phenom. \textbf{57}, 243 (1998).
\bibitem{fsm}
P. Aebi et al., Phys. Rev. Lett. \textbf{72}, 2757 (1994); T.~J. Kreutz et al., Phys. Rev. B 
\textbf{58}, 1300 (1998); Th. Straub et al., Phys. Rev. B \textbf{55}, 13473 (1997).
\bibitem{osterwalder}
J. Osterwalder, T. Greber, S. H\"ufner, L. Schlapbach, Phys. Rev. B \textbf{44}, 13764 (1991).
\bibitem{norm}
Normalization is achieved by dividing intensities on each azimuthal 
scan, i.e. concentric circle around $\Gamma$, by its mean intensity.
\bibitem{calc}
A.~M. Woolley, G. Wexler, J. Phys. C \textbf{10}, 2601 (1977); 
H.~W. Myron, A.~J. Freeman, Phys. Rev. B \textbf{11}, 2735 (1975).
\bibitem{matsushita}
T. Matsushita, S. Imada, H. Daimon, T. Okuda, K. Yamaguchi, H. Miyagi, 
S. Suga, Phys. Rev. \textbf{B} 56, 7687 (1997).
\bibitem{claessen90}
R. Claessen, B. Burandt, H. Carstensen, M. Skibowski, Phys. Rev. B \textbf{41}, 8270 (1990).
\bibitem{pillo99}
Th. Pillo et al., to be published
%\bibitem{campuzano96}
%J.÷C. Campuzano, H. Ding, M.÷R. Norman, M. Randeria, A.÷F. Bellman, T. 
%Yokoya, T. Takahashi, H. Katayama-Yoshida, T. Mochiku, K. Kadowaki, 
%Phys. Rev. B \textbf{53}, R14737 (1996).
\bibitem{pillo99b}
Th. Pillo, J. Hayoz, P. Aebi, L. Schlapbach, J. Electron Spectr. Relat. Phenom., in press (1999).
%\bibitem{geertsma72}
%W. Geertsma, C. Haas, R. Huisman, F. Jellinek, Sol. State Comm. \textbf{10}, 75 (1972).
%\bibitem{mott90}
%N.~F. Mott: \textit{Metal-Insulator Transitions}, Taylor \& Francis 
%(1990) p. 138;~F. Gebhard: \textit{The Mott Metal-Insulator Transition}, 
%Springer Tracts in Modern Physics 137, Springer (1997) p. 34.

%\bibitem{wilson69}
%J.~A. Wilson, A.~D. Yoffe, Adv. Phys. 18 (1969) 193.%
%\bibitem{claessen92}
%R. Claessen \underline{et al.}, Phys. Rev. Lett. 69 (1992) 808; R. 
%Claessen \underline{et al.}, Physica B 230-232 (1997) 294
%\bibitem{anderson85}
%O. Anderson \underline{et al.}, Phys. Rev. Lett. 55 (1985) 2188 
%and references and therein
%\bibitem{withers86}
%R.~L. Withers, J.~A. Wilson, J. Phys. C 19 (1986) 4809.
%\bibitem{scruby75}
%C.~B. Scruby, P.~M. Williams, G.~S. Parry, Phil. Mag. 31 (1975) 255.
%\bibitem{spijkerman97}
%A. Spijkerman \textit{et al.}, Phys. Rev. B 56 (1997) 13757.
%\bibitem{pollak81}
%R.~A. Pollak \textit{et al.}, Phys. Rev. B 24 (1981) 7435.

%\bibitem{okuda98}
%T. Okuda \textit{et al.}, J. Electron Spectrosc. Relat. Phenom. 
%88-91 (1998) 287.
%\bibitem{woolley77}
%A.~M. Woolley, G. Wexler, J. Phys. C 10 (1977) 2601, and references therein.
%\bibitem{myron75}
%H.~W. Myron, A.~J. Freeman, Phys. Rev. B 11 (1975) 2735.
%\bibitem{aebi98}
%P. Aebi \textit{et al.}, Surf. Sci. 402-404 (1998) 614 and references therein.
%\bibitem{smith75}
%N.~V. Smith, M.~M. Traum, Phys. Rev. B 11 (1975) 2078.


\end{thebibliography}

%%%%%%%%%%%%%%%%%%%%%%%%%%%%%%%%%%%%%%%%%%%%%%%%%%%%%%%%%%%%%%%%%%%%%%%%%

\begin{figure}
	%\centerline{\epsfig{file=fig1.epsf,width=8.6cm}}
	\vspace{0.3cm}
	\caption{Fermi surface mapping results of 1$T$-TaS$_{2}$, taken with monochromatized
	He-I$\alpha$ radiation (21.2 eV). Map (a) has
	been obtained at room temperature, i.e., in the quasicommensurate CDW phase.
	Map (b) was measured at 140~K, in the commensurate CDW phase. The 
	maps have been obtained by a simple renomalization process taking into account the space group
	symmetry to enhance outerlying weak features (see text).}
	\label{fsm}
\end{figure}
\begin{figure}
	%\centerline{\epsfig{file=fig2.epsf,width=8cm}}
	\vspace{0.3cm}
	\caption{ARPES spectra along the Fermi surface contour, performed at 
	room temperature (left side) and at 140÷K (right side). Enumeration
	of spectra is shown at the right by the arrow, corresponding to the 
	drawing in Fig.~3. Raw data is shown without any normalization. The 
	two quasiparticle peaks A and B are explained in the text.}
	\label{contour}
\end{figure}
\begin{figure}
	%\centerline{\epsfig{file=fig3.epsf,width=8cm}}
	\vspace{0.3cm}
	\caption{Simplifiying sketch of the band dispersion and the Fermi 
	surface in 1$T$-TaS$_2$. The upper panel mimicks the 
	E(\textbf{k}) dispersion of the unperturbed Ta 5$d$ band in 
	the $\Gamma M$ azimuthal direction. The CDW induced subbands are 
	indicated by the shady areas. The lower panel sketches the 
	inverse situation, the \textbf{k$_{\|}$}(E=E$_F$) dispersion, i.e., the 
	Fermi surface, built up by the elliptic shape of the Ta 5$d$ band. 
	The arc along the FS ellipse denotes the location in 
	\textbf{k}-space where the ARPES spectra of Fig.~2 have been measured.
	The two circular arcs show where the ARPES spectra of Fig.~4 have 
	been measured.}
	\label{sketch}
\end{figure}
\begin{figure}
	%\centerline{\epsfig{file=fig4a.epsf,width=8cm}}
	%\centerline{\epsfig{file=fig4b.epsf,width=8cm}}
	\caption[last]{(a) Room temperature ARPES spectra measured along the circular arcs of Fig.~3, i.e., 
	with two Fermi surface crossings, for a polar angle of 
	$\theta=32^\circ$. Location in \textbf{k}-space is 
	indicated at the bottom, meaning spectra are from $\left [\Gamma 
	\mbox{K} \right ]$ over $\left [\Gamma \mbox{M} \right ]$ to $\left [\Gamma 
	\mbox{K} \right ]$. The arrows indicate the Fermi 
	surface crossings from  the n(\textbf{k}) analysis (see text). In 
	the spectra a$\rightarrow$b on the right the dispersing 
	quasiparticle peak is indicated by small ticks.
	(b) ARPES spectra as in (a), but at $\theta=50^\circ$ polar angle, i.e., 
	outside the Peierls gap region (gray region on the 
	ellipse). As in (a), spectra are shown in the right panel from c to d, as 
	indicated in the gray scale plot and in 
	Fig.÷3}
	\label{azi}
\end{figure}
\end{multicols}
\end{document}